# Light-induced inverse spin Hall effect and field-induced circular photogalvanic effect in GaAs revealed by two-dimensional terahertz Fourier analysis

Tomohiro Fujimoto,[*] Yuta Murotani, Tomohiro Tamaya, Takayuki Kurihara[†], Natsuki Kanda[‡], Changsu Kim, Jun Yoshinobu, Hidefumi Akiyama, Takeo Kato, and Ryusuke Matsunaga[§]

The Institute for Solid State Physics, The University of Tokyo, Kashiwa, Chiba 277-8581, Japan

[*]fujimoto@issp.u-tokyo.ac.jp
[§]matsunaga@issp.u-tokyo.ac.jp
[†]Present address: Department of Basic Science, Graduate School of Arts and Sciences, The University of Tokyo, 3-8-1 Komaba, Meguro-ku, Tokyo 153-8902, Japan.
[‡]Present address: RIKEN Center for Advanced Photonics, RIKEN, 2-1 Hirosawa, Wako, Saitama 351-0198, Japan

**Abstract**.

The electromotive force transverse to a bias field under irradiation of circularly polarized light, namely the photovoltaic Hall response or light-induced anomalous Hall effect, has attracted considerable attention to investigate the topologically nontrivial states in Floquet engineering and the inverse spin Hall effect of spin-polarized carriers in spintronics. However, taking into account inversion symmetry breaking by the bias field, the circularly polarized light can excite photocarriers with asymmetric momentum distribution, which generates injection current transverse to the bias field. Therefore, the field-induced circular photogalvanic effect (FI-CPGE) should also emerge in the very same experimental configuration for light-induced anomalous Hall effect but has been overlooked in literature. In this work, using terahertz pulses as a bias field for a semiconductor GaAs, we conduct two-dimensional Fourier analysis and demonstrate that FI-CPGE can play a major role in the photovoltaic Hall response. Counterintuitively, FI-CPGE is significantly enhanced when the photocarriers are excited near the bandgap with small density of states and low group velocity, which can be explained by a three-level resonant nonlinear interaction near the band degeneracy point. We also clarified that FI-CPGE would be further largely detected in the contact-type measurement using electrodes because of the absence of a filtering effect inherent to terahertz pulses. This work provides a comprehensive, generalized view of the photovoltaic Hall response in biased materials, paving a new avenue for detecting topological monopoles in momentum space hidden in equilibrium using third-order nonlinear responses.

**Main text**

Nonthermal control of matter using intense light fields has recently attracted tremendous interest in ultrafast optical science and condensed matter physics because nontrivial states can manifest themselves in nonequilibrium [1–3]. Of particular interest is the case driven by a circularly polarized light (CPL); based on the Floquet theory, a periodic CPL field can be embedded in light-dressed states of electrons with an artificial gauge field and the associated Berry curvature [4]. The topologically nontrivial states of electrons under light illumination are expected to produce an anomalous Hall current to a bias field [5–10]. Beyond simple CPL, counterrotating bicircular light pulses [11–13] have also attracted growing interest as a way to impose a specific rotational symmetry on matter, which may also produce nontrivial states with finite Hall conductivity [14–16].

However, interpretation of the transverse current observed under CPL irradiation in biased materials requires special care. A representative example is the light-induced anomalous Hall effect in a two-dimensional (2D) Dirac system, which is theoretically anticipated to be a signature of the Floquet-topological insulating phase [5]. Although the light-induced transverse current in biased graphene was recently observed [17], theoretical analysis revealed that the Floquet-topological phase plays only a minor role in the current, and the current is dominated by the momentum asymmetry of photoexcited carriers [18], as schematically shown in Fig. 1(a). This process can be understood as a nominally second-order photocurrent generation by CPL due to broken inversion symmetry by dc bias field, which we term the field-induced circular photogalvanic effect (FI-CPGE) [19]. As long as CPL and bias field are applied to a material simultaneously, both the Hall current due to time-reversal symmetry breaking by CPL and the photocurrent due to inversion symmetry breaking by bias field should contribute to the transverse current on the same footing [19]. Although most theoretical studies of Floquet engineering and light-field control of matter have considered only the former mechanism, a recent experiment has also demonstrated that the latter, FI-CPGE, is dominant in a three-dimensional Dirac semimetal [19]. Therefore, the next issue to be resolved for Floquet engineering is how FI-CPGE appears in more generalized systems and when it can be suppressed. Notably, FI-CPGE, a kind of helicity-dependent photovoltaic Hall responses, has attracted much theoretical attention itself [20–25] from diverse viewpoints including Riemannian geometry [20]. Despite its fundamental aspect, the microscopic mechanism of the photovoltaic Hall current is complicated and highly nontrivial due to the coexistence of light, bias, and their correlation. Interestingly, given the topological aspect of CPGE described by the Berry curvature in non-centrosymmetric crystals [26,27], FI-CPGE may be viewed as implementation of an effective Berry curvature in solids using bias field. Therefore, comprehensive understanding of FI-CPGE is demanded for exploring a new state of matter with novel functionality for large second-order responses.

To capture a general perspective for interactions with light and bias fields, a semiconductor GaAs offers a good platform owing to the well-defined bandgap and multiple valence bands involved in optical transitions, in contrast to the simple Dirac systems. The band touching point between the light-hole (LH) and heavy-hole (HH) bands at the zone center hosts doubly degenerated monopoles in momentum space with divergent Berry curvatures and opposite signs [28]. Therefore, although the Floquet-topological phase transition predicted in gapless systems can be ignored, intriguing responses related to band topology would manifest themselves when its symmetry is broken by perturbation. Indeed, spin-polarized carriers excited by CPL flow into a transverse direction to a bias field owing to the Berry curvature, as shown in Fig. 1(b), which we term the light-induced inverse spin Hall effect (LI-ISHE). LI-ISHE has been studied in the field of spintronics using a static bias field and continuous-wave (CW) light [29–32]; to the best of our knowledge, however, the contribution of FI-CPGE has been neglected in literature for spintronics. To discriminate between FI-CPGE and LI-ISHE, time-resolved detection of the transient Hall conductivity *during* and *after* CPL pump pulse irradiation is pivotal because FI-CPGE occurs only *during* the pump, whereas LI-ISHE occurs even *after* the pump as long as the spin polarization survives [19, 33]. Using optical pump-terahertz (THz) probe spectroscopy, the ISHE of electrons in GaAs was observed *after* the pump, realizing a quantitative evaluation of the spin Hall conductivity spectrum [34]. The ISHE of holes has also been observed in a similar way in GaAs quantum wells at low temperature [35]. However, the transverse current *during* CPL irradiation in conventional semiconductors has not been investigated, despite its practical importance in more common situations using static bias and CW light. Although the previous study of a Dirac semimetal could elucidate the dominance of FI-CPGE based on the sign of $\sigma_{yx}(\omega)$ [19], such a qualitative factor can be used only in simple systems described by a 2-band model, and therefore it is not directly applicable to other semiconductors. A new strategy for the quantitative evaluation of FI-CPGE and LI-ISHE must be developed for deeper insight into the fundamental interaction between light, matter, and bias field, which will further exploit novel functionalities related to band topology for spintronics as well as for Floquet engineering.

In this Letter, we conduct optical pump-THz Faraday probe spectroscopy to investigate the transient photovoltaic Hall response in bulk semiconductor GaAs *during* irradiation with near-infrared (NIR) CPL pulses. Our 2D frequency-domain analysis successfully discriminates between LI-ISHE and FI-CPGE. Even though the detection of FI-CPGE is significantly suppressed by the spectral filter inherent to THz spectroscopy, we show that FI-CPGE plays a major role in the observed signal with remarkable enhancement near the bandgap. The results agree well with the microscopic theory, unveiling a drastic role of three-revel resonance of FI-CPGE in the multiple energy bands in contrast to the simple Dirac systems. This work presents a comprehensive understanding of the origins of the CPL-induced anomalous Hall current.

For sample preparation, we first grew 1.0 μm-thick non-doped GaAs (001) sandwiched between 1.7 μm-thick $Al_{0.19}Ga_{0.81}As$ protective layers using the molecular-beam epitaxy method on a non-doped GaAs substrate. The substrate was then removed by mechanical polishing and selective chemical etching over a diameter of 4 mm to make a free-standing sample suitable for transmission measurements. All the experiments were performed at room temperature, where the bandgap $E_g$ is 1.42 eV. Figure 1(c) shows the experimental setup. The output of a Yb:KGW laser amplifier was separated into three beams. One was converted into NIR CPL pulses with a tunable pump photon energy $\hbar\Omega_{\text{pump}}$ between 1.38 and 1.56 eV using an optical parametric amplifier. The other two beams were used for THz pulse generation and for the gate pulses to detect the THz pulses in a GaP crystal. The incident THz pulses were linearly polarized in the $x$ direction. We used a pair of wire-grid polarizers for the transmitted THz pulses to detect the $y$ component originating from the transverse current in the sample, depending on the pump helicity [34]. The delay times $t_1$ and $t_2$ were scanned by changing the beam path lengths of the THz probe and NIR pump pulses, respectively. To extract the signal depending on the pump helicity, we evaluated the $y$ component as the difference between the signals for left- and right-handed CPL. More detailed information is available in Supplemental Material [36].

The upper panel of Fig. 2(a) shows $E_x(t_1)$, the THz probe pulse waveform after transmitting the sample without a pump. The middle panel shows 2D plots of $E_y(t_1, t_2)$, the Faraday rotation signal induced by the CPL pump. Here, $\hbar\Omega_{\text{pump}}$ is 1.46 eV and the pump fluence is 16 µJ cm$^{-2}$, for which the $E_y$ signal oscillates along the $t_1$-axis. The bottom panel shows the data for a larger pump fluence of 210 µJ cm$^{-2}$. Remarkably, a distinct diagonally oscillating waveform was observed. The qualitative differences in the 2D plots between the weak and strong excitations suggest that the main mechanism of the CPL-induced THz Faraday rotation changes with pump fluence.

The 2D data in the middle panel of Fig. 2(a) is similar to the recently reported LI-ISHE in GaAs [34]. From the fitting analysis [36], the $E_y$ signal decays along the $t_2$-axis with fast (220 fs) and slow (82 ps) components, which coincide well with the spin relaxation times of photoexcited holes [44] and electrons [45], respectively. Based on spectral analysis, the slowly decaying smaller signal has been well explained by LI-ISHE of spin-polarized electrons [34]. Similarly, the fast-decaying larger signal at $t_2$ = 0 ps in the middle panel is attributed to LI-ISHE of short-lived spin-polarized holes with stronger spin-orbit interaction [35,47]. By contrast, the diagonally oscillating signal in the bottom panel of Fig. 2(a) is similar to that of the recently reported FI-CPGE in a Dirac semimetal [19]. Therefore, the data suggests that the observed Faraday rotation signal may be accounted for by LI-ISHE (FI-CPGE) in the weak (strong) excitation regime, as examined below.

To discuss LI-ISHE of holes and FI-CPGE, which temporally overlap in $E_y(t_1, t_2)$, we conducted a 2D-Fourier transform and plotted $|E_y(\omega_1, \omega_2)|$ in Fig. 2(b). In the 2D frequency domain, a single tilted ellipse at $\omega_2/2\pi \sim 0$ THz appears for the weaker excitation (left panel), while a second ellipse at $\omega_2/2\pi \sim -2.5$ THz appears for the stronger excitation (right panel). To understand these features, the nonlinear response functions of both processes were considered. In LI-ISHE, the pump pulse leaves spin-polarized carriers that respond to the bias field $E_x(t_1)$ with a finite Hall conductivity and decay along $t_2$ owing to spin relaxation. By contrast, FI-CPGE generates a photocurrent only at the pump arrival under the bias field, i.e., at $t_2 = t_1$. Therefore, each response function can be modeled as

$$\sigma^{(3)}_{LI-ISHE}(t_1, t_2) = \Theta(t_2)e^{-\gamma_s t_2} \times \sigma^{yx}_{ISHE}(t_1) \quad (1)$$

$$\sigma^{(3)}_{FI-CPGE}(t_1, t_2) = \Theta(t_2)e^{-\Gamma t_2} \times \alpha_{CPGE}\delta(t_2 - t_1), \quad (2)$$

where $\Theta(t)$ is a step function, $\delta(t)$ is a delta function, and $\Gamma$ and $\gamma_s$ are the decay constants. $\sigma_{ISHE}(t_1)$ represents the inverse spin Hall conductivity of holes. $\alpha_{CPGE}$ represents the efficiency of FI-CPGE. The last two quantities can be highly dependent on $\hbar\Omega_{pump}$, as discussed later. The essential difference between the two response functions is evident in the 2D frequency domain:

$$\sigma^{(3)}_{LI-ISHE}(\omega_1, \omega_2) = \sigma^{yx}_{ISHE}(\omega_1)\frac{\gamma_s}{\gamma_s - i\omega_2}, \quad (3)$$

$$\sigma^{(3)}_{FI-CPGE}(\omega_1, \omega_2) = \frac{\Gamma\alpha_{CPGE}}{\Gamma - i(\omega_1 + \omega_2)}. \quad (4)$$

Assuming that $\sigma^{yx}_{ISHE}(\omega_1)$ and $\alpha_{CPGE}$ are constant for simplicity, Eqs. (3) and (4) are schematically shown in Figs. 3(a) and 3(b), respectively. The responses of LI-ISHE and FI-CPGE peak at $\omega_2 = 0$ and $\omega_1 + \omega_2 = 0$, respectively. Note that the observed signal $E_y(\omega_1, \omega_2)$ in the experiment is not solely determined by the nonlinear conductivity in Eq. (3) or (4) because the detectable spectral region can be restricted by other extrinsic factors: the probe pulse spectrum $E_x^{bias}(\omega_1)$, pump pulse envelope spectrum $I^{pump}(\omega_2)$, and propagation function $G(\omega_1 + \omega_2)$ that relates the transverse current $J_y$ to the detected electric field $E_y$ (see Fig. S1 and S5 in Supplemental Material [36]). Importantly, $G(\omega_1 + \omega_2)$ necessarily produces a notch downward to the right along the line of $\omega_1 + \omega_2 = 0$, where the detection efficiency is precisely zero. This is a consequence of diffraction; the DC-limit electromagnetic wave does not propagate in free space and thus never detected in noncontact THz spectroscopy [46]. Figure 3(c) summarizes the available spectral region, where two ellipses appear as a result of filtering. Considering these extrinsic filters, we phenomenologically calculated the expected signals in the 2D temporal and frequency domains originating from LI-ISHE and FI-CPGE [36], as plotted in Figs. 3(d) and 3(e), respectively, which reproduce the experimental data in Figs. 2(a) and 2(b) very well. This confirmed that the observed signal of the CPL-induced THz Faraday rotation was well explained by LI-ISHE (FI-CPGE) at small (large) pump fluence.

It should be emphasized that the two ellipses observed in the stronger pump in Fig. 2(b) are just the "tails" of FI-CPGE. The peak of FI-CPGE exists along $\omega_1 + \omega_2 = 0$, but the propagation function $G(\omega_1 + \omega_2)$ prevents this main part from detection. Nevertheless, the experiment showed that the small tails of FI-CPGE can dominate the THz signal under certain conditions, which reveals the significance of FI-CPGE, although this has been disregarded in many studies. FI-CPGE signal would appear much more efficiently in contact-type measurements such as conventional DC transport and on-chip THz spectroscopy [17]. In other words, THz spectroscopy is suitable for studying LI-ISHE with efficiently suppressing FI-CPGE.

While the lower ellipse at $\omega_2/2\pi \sim -2.5$ THz is solely attributed to FI-CPGE, the upper ellipse at $\omega_2/2\pi \sim 0$ THz includes both LI-ISHE and FI-CPGE signals. To fully understand the dependence on pump fluence, we integrated the two ellipse signals separately and plotted the current amplitude $A_y$ as a function of pump fluence, as shown in Fig. 3(f) (see also Supplemental Material [36]). In contrast to the simple linearity of the lower ellipse, the upper ellipse saturates in the moderate fluence region (50–100 µJ cm$^{-2}$) and increases again for larger fluences (>100 µJ cm$^{-2}$). This nonmonotonic behavior is consistent with the main contribution of LI-ISHE (FI-CPGE) for weaker (stronger) excitation. LI-ISHE originates from the Hall conductivity of photoexcited carriers but is accompanied by the Drude response, which suppresses the penetration of the incident THz wave as the carrier density increases for stronger excitation. The dashed line in Fig. 3(f) shows the modeled fluence dependence of LI-ISHE accounting for this suppression [36]. By contrast, FI-CPGE is a field-assisted generation of photocurrent, which is relatively robust against pump fluence unless the interband absorption saturates. Therefore, the upper ellipse signal in the strong excitation regime was attributed to FI-CPGE, even though the main part of FI-CPGE was filtered out.

Finally, we discuss the dependence of LI-ISHE and FI-CPGE on pump photon energy $\hbar\Omega_{\text{pump}}$. Figures 4(a) and 4(b) show the upper and lower ellipse signals, $A_y^{\text{Upper}}$ and $A_y^{\text{Lower}}$, respectively, with a fixed fluence of 100 μJ cm$^{-2}$. Both signals show a similar enhancement near the bandgap. While $A_y^{\text{Lower}}$ in Fig. 4(b) is solely attributed to FI-CPGE, $A_y^{\text{Upper}}$ in Fig. 4(a) includes LI-ISHE and FI-CPGE, and the peak is notably shifted compared with Fig. 4(b). The result indicates that the dominant mechanism is different, *i.e.*, the peak in Fig. 4(a) is attributed to LI-ISHE. We calculated $\sigma^{(3)}_{\text{LI-ISHE}}$ and $\sigma^{(3)}_{\text{FI-CPGE}}$ from the microscopic model based on the 6-band Kane Hamiltonian [36], according to the third-order susceptibility derived in a pioneering work [25]. The results of our calculations for GaAs in Figs. 4(a) and 4(b) agree very well with the experimental data, including the different peak positions. This result can be explained by each microscopic origin. LI-ISHE is driven by the Berry curvature of photoexcited holes [47], which is divergently enhanced at the valence band maximum where LH and HH bands touch as shown in Fig. 4(c) [28]. By contrast, FI-CPGE originates from the third-order nonlinear interaction involving interband transitions between LH, HH, and conduction bands (CB), as shown in Fig. 4(d). In this work, we used the THz probe pulse centered at $\omega_{\text{probe}}/2\pi=1.2$ THz, which resonates with the LH–HH transition at finite momenta near the Γ-point. Then, a pump pulse with $\hbar\Omega_{\text{pump}}$ slightly higher than the bandgap can make all three optical transitions in Fig. 4(d) resonant to the real states of LH, HH, and CB without resorting to virtual states. Thus, FI-CPGE is resonantly enhanced slightly above the band degeneracy point, depending on the bias field frequency $\omega_{\text{probe}}$ and the CB dispersion relation. In Supplemental Material, the result is compared to the case of a single valence band [36], which reveals that the three-level resonance greatly enhances FI-CPGE by several tens of times near the band gap. The remarkable enhancement of FI-CPGE has important implications. Because FI-CPGE is described by an asymmetric momentum distribution of photocarriers [18], one might intuitively assume that FI-CPGE should be reduced near the band edge because of less joint density of states and vanishingly small group velocity. However, this study clearly shows that FI-CPGE is resonantly enhanced toward the band edge as long as the nearly degenerate bands are involved in the optical transitions. This result suggests that FI-CPGE plays an even more important role in the light-induced anomalous Hall effect when multiple energy bands are involved with optical transition.

In summary, we demonstrated that the microscopic origins of the CPL-induced anomalous Hall response in GaAs can be resolved by 2D Fourier analysis. Although the spectral filtering suppresses the detection efficiency of FI-CPGE, it appears very strongly when the nearly degenerate bands are involved in the optical transition. The results shed light on the interpretation of the CPL-induced anomalous Hall effect and photovoltaic Hall response. Because LI-ISHE and FI-CPGE are sensitive to the band degeneracy and dispersion around it, they can be used to identify the monopole charge hidden in equilibrium. Moreover, FI-CPGE can be selectively suppressed by tuning the pump frequency away from interband transitions associated with degenerated bands or by using intense field far beyond the perturbative regime, which will help the targeted observation of contributions anticipated in Floquet engineering. Furthermore, the significant magnitude of FI-CPGE substantiates that CPGE can be efficiently manipulated by THz bias field. The artificial controllability of the topological property using bias field may be a novel counterpart to the CPL-induced Berry curvature in Floquet engineering [5], as well as to the control of Berry curvature and Berry connection in the strain engineering [48-50]. Further progress in theoretical description of light-matter-bias interaction will exploit rich playground to realize topologically non-trivial states of matter with giant responses.

**Acknowledgements**

This work was supported by JST FOREST (Grant No. JPMJFR2240), JST PRESTO (Grants No. JPMJPR2006, and No. JPMJPR2107), JST CREST (Grant No. JPMJCR20R4), and JSPS KAKENHI (Grants No. JP24K00550, and No. JP24K16988, and No. JP20K22478). R.M. acknowledges support from MEXT Quantum Leap Flagship Program (MEXT Q-LEAP, Grant No. JPMXS0118068681). R.M. and T.F. conceived the project. C.K., T.F., and H.A. fabricated the sample. T.F., Y.M., and N.K. developed the pump-probe spectroscopy system with the help of T.Kurihara, J.Y., and R.M. T.F. performed the experiment and analyzed the data with help of Y.M. T.F. conducted theoretical calculation with help of Y.M, T.T., and T.Kato. All the authors discussed the results. T.F. and R.M. wrote the manuscript with substantial feedback from Y.M. and all the co-authors.

**References**

1. A. de la Torre, D. M. Kennes, M. Claassen, S. Gerber, J. W. McIver, and M. A. Sentef, Colloquium: Nonthermal pathways to ultrafast control in quantum materials. Rev. Mod. Phys. **93**, 041002 (2021).
2. C. Bao, P. Tang, D. Sun, and Shuyun Zhou, Light-induced emergent phenomena in 2D materials and topological materials. Nat. Rev. Phys. **4**, 33 (2021).
3. M. Borsch, M. Meierhofer, R. Huber, and M. Kira, Lightwave electronics in condensed matter. Nat. Rev. Mater. **8**, 668 (2023).
4. T. Oka and S. Kitamura, Floquet Engineering of Quantum Materials. Annu. Rev. Condens. Matter Phys. **10**, 387 (2019).
5. T. Oka and H. Aoki, Photovoltaic Hall effect in graphene. Phys. Rev. B **79**, 081406(R) (2009).
6. X. Wang, E. Ronca, and M. A. Sentef, Cavity quantum electrodynamical Chern insulator: Towards light-induced quantized anomalous Hall effect in graphene. Phys. Rev. B **99**, 235156 (2019).
7. M. Nuske, L. Broers, B. Schulte, G. Jotzu, S. A. Sato, A. Cavalleri, A. Rubio, J. W. McIver, and L. Mathey, Floquet dynamics in light-driven solids. Phys. Rev. Research **2**, 043408 (2020).
8. P. X. Nguyen and W.-K. Tse, Photoinduced anomalous Hall effect in two-dimensional transition metal dichalcogenides. Phys. Rev. B **103**, 125420 (2021).
9. Y. Hirai, S. Okumura, N. Yoshikawa, T. Oka, and R. Shimano, Floquet Weyl states at one-photon resonance: An origin of nonperturbative optical responses in three-dimensional materials. Phys. Rev. Research **6**, L012027 (2024).
10. N. Yoshikawa, S. Okumura, Y. Hirai, K. Ogawa, K. Fujiwara, J. Ikeda, A. Ozawa, T. Koretsune, R. Arita, A. Mitra, A. Tsukazaki, T. Oka, and R. Shimano, Light-induced anomalous Hall conductivity in massive 3D Dirac semimetal $Co_3Sn_2S_2$, arXiv:2209.11932v2.
11. I. Tyulnev, Á. Jiménez-Galán, J. Poborska, L. Vamos, P. St. J. Russell, F. Tani, O. Smirnova, M. Ivanov, R. E. F. Silva, and J. Biegert, Valleytronics in bulk $MoS_2$ with a topologic optical field. Nature **628**, 746 (2024).
12. S. Mitra, Á. Jiménez-Galán, M. Aulich, M. Neuhaus, R. E. F. Silva, V. Pervak, M. F. Kling, and S. Biswas, Light-wave-controlled Haldane model in monolayer hexagonal boron nitride. Nature **628**, 752 (2024).
10

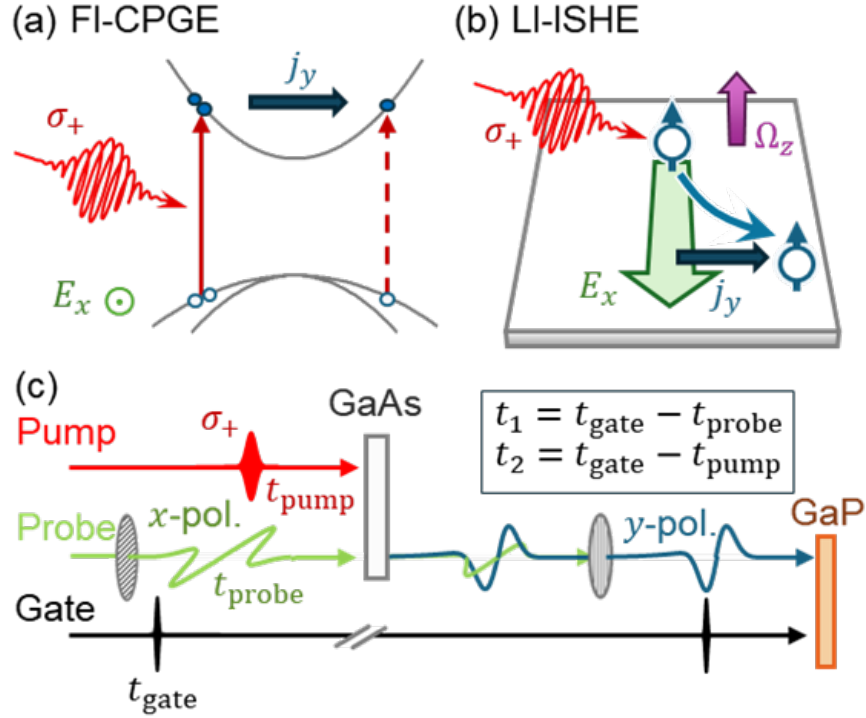

FIG. 1. (a)(b) Origins of the light-induced anomalous Hall effect or photovoltaic Hall response in semiconductors: FI-CPGE due to the momentum asymmetry of photoexcited carriers and LI-ISHE driven by the Berry curvature. (c) Configuration of circularly polarized NIR pump-THz Faraday probe spectroscopy. $t_1$ and $t_2$ are scanned by changing the optical beam path lengths of the THz probe and NIR pump pulses, respectively.

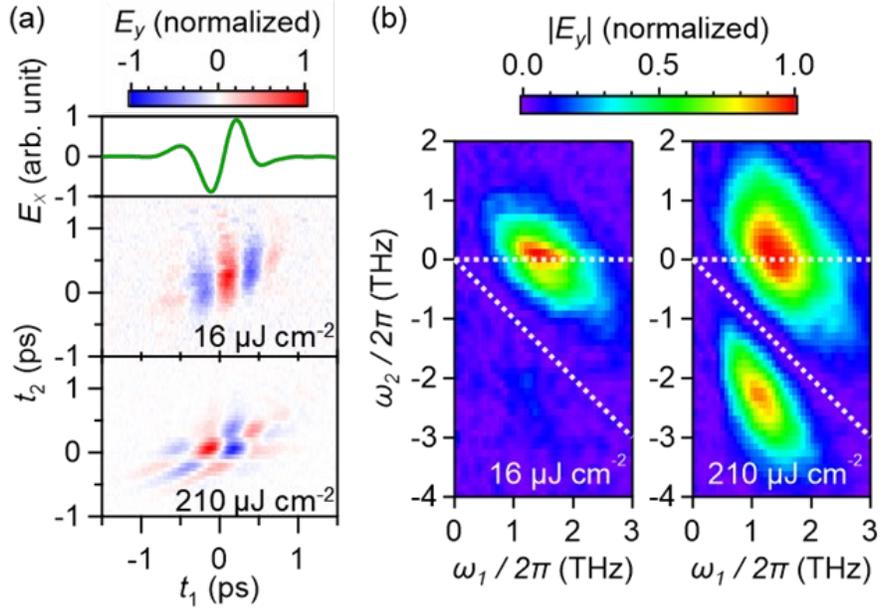

FIG. 2. (a) THz probe pulse waveform $E_x(t_1)$ after transmission (top). 2D plots of THz Faraday rotation signal $E_y(t_1, t_2)$ induced by CPL for fluences of 16 μJ cm$^{-2}$ (center) and 210 μJ cm$^{-2}$ (bottom). (b) 2D Fourier plots of the Faraday rotation signal $|E_y(\omega_1, \omega_2)|$ for fluences of 16 μJ cm$^{-2}$ (left) and 210 μJ cm$^{-2}$ (right). The horizontal (diagonal) dotted lines correspond to $\omega_2 = 0$ ($\omega_1 + \omega_2 = 0$).

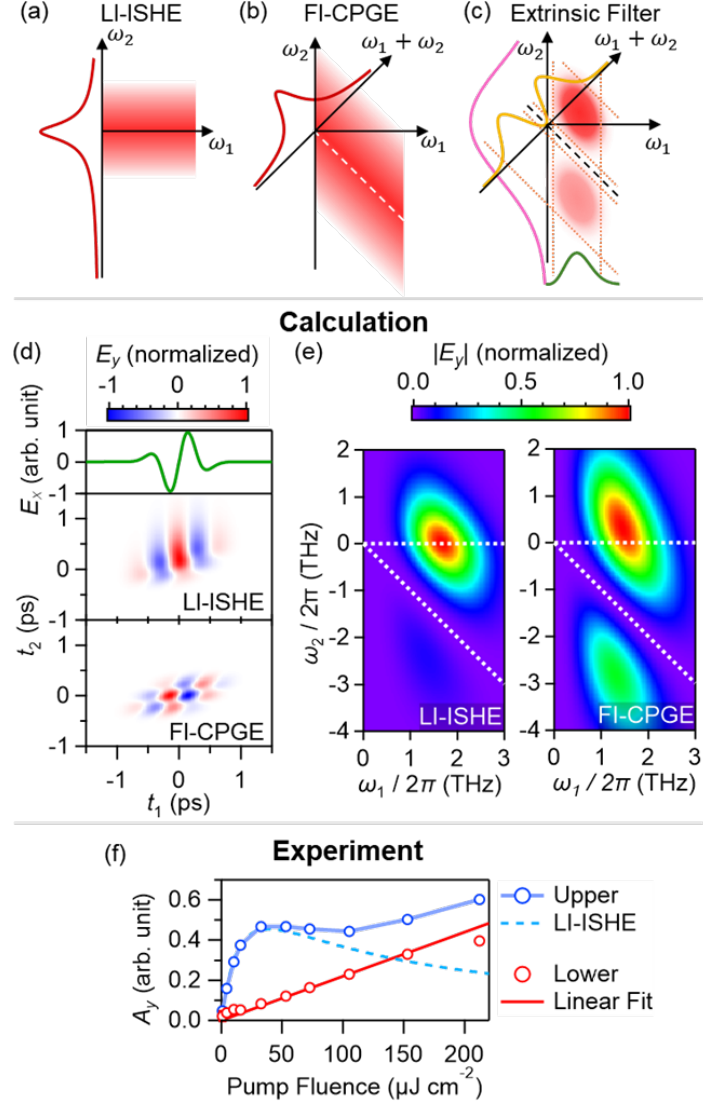

FIG. 3. 2D Response functions of (a) LI-ISHE and (b) FI-CPGE modeled using Eqs. (3) and (4), respectively. (c) Extrinsic filters limiting the detectable spectral regions in optical pump-THz probe spectroscopy: probe spectrum (green), pump envelope spectrum (pink), and propagation function (orange). The diagonal dashed lines in (b) and (c) correspond to $\omega_1 + \omega_2 = 0$. (d) THz probe pulse waveform $E_x(t_1)$ used for the calculation (top). 2D plots of the calculated Faraday rotation signal $E_y(t_1, t_2)$ for LI-ISHE (middle) and FI-CPGE (bottom). (e) 2D Fourier plots of the Faraday rotation signal $|E_y(\omega_1, \omega_2)|$ for LI-ISHE (left) and FI-CPGE (right). The horizontal (diagonal) dotted lines correspond to $\omega_2 = 0$ ($\omega_1 + \omega_2 = 0$). (f) Pump fluence dependence of the current amplitude for the upper and lower ellipses in the 2D Fourier plots in the experiment.

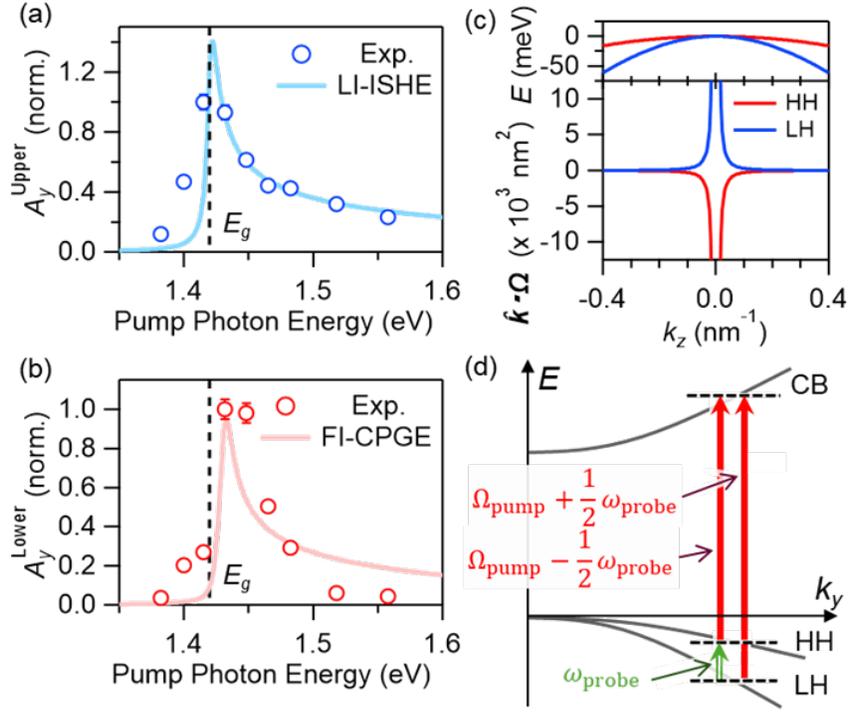

FIG. 4. Pump photon energy dependence of the current amplitudes for (a) upper and (b) lower ellipses with a fixed fluence of 100 μJ cm$^{-2}$, which are attributed to LI-ISHE and FI-CPGE, respectively. The markers and curves show the results of experiments and calculations, respectively. (c) Calculated band structure near the band edge in GaAs (top) and the Berry curvature (bottom). (d) Diagram of nonlinear interactions involving LH, HH, and CB.